\documentclass[apj]{emulateapj}

\begin{document}
\title{The Broadband Infrared Emission Spectrum of the Exoplanet TrES-3.}
\author{Francois Fressin\altaffilmark{1,2}, Heather A. Knutson\altaffilmark{1}, David Charbonneau\altaffilmark{1}, Francis T. O'Donovan\altaffilmark{1}, Adam Burrows\altaffilmark{3}, Drake Deming\altaffilmark{4}, Georgi Mandushev\altaffilmark{5} and David Spiegel\altaffilmark{3}}
\altaffiltext{1}{Harvard-Smithsonian Center for Astrophysics, 60 Garden St., Cambridge, MA 02138}
\altaffiltext{2}{ffressin@cfa.harvard.edu}
\altaffiltext{3}{Department of Astrophysical Sciences, Peyton Hall Rm. 105, Princeton, NJ 08544-1001}
\altaffiltext{4}{NASA Goddard Space Flight Center, 8800 Greenbelt Road, Greenbelt, Maryland 20771}
\altaffiltext{5}{Lowell Observatory, 1400 West Mars Hill Road, Flagstaff, AZ 86001}

\begin{abstract}

We use the \emph{Spitzer Space Telescope} to estimate the dayside thermal emission of the exoplanet TrES-3 integrated in the 3.6, 4.5, 5.8, and 8.0~\micron~bandpasses of the Infrared Array Camera (IRAC) instrument.  We observe two secondary eclipses and find relative eclipse depths of $0.00346\pm0.00035$, $0.00372\pm0.00054$, $0.00449\pm0.00097$, and $0.00475\pm0.00046$, respectively in the 4 IRAC bandpasses. We combine our results with the earlier $K$ band measurement of De Mooij et al. (2009), and compare them with models of the planetary emission. We find that the planet does not require the presence of an inversion layer in the high atmosphere. This is the first very strongly irradiated planet that does not have a temperature inversion, which indicates that stellar or planetary characteristics other than temperature have an important impact on temperature inversion. 
De Mooij \& Snellen (2009) also detected a possible slight offset in the timing of the secondary eclipse in $K$ band.
However, based on our 4 \emph{Spitzer} channels, we place a $3 \sigma$ upper limit of $|e cos(\omega)| < 0.0056$ where \emph{e} is the planet's orbital eccentricity and \emph{$\omega$} is the longitude of the periastron. This result strongly indicates that the orbit is circular, as expected from tidal circularization theory.                                                    
\end{abstract}

\keywords{planetary systems - stars: individual: TrES-3 - techniques: photometric - eclipses - infrared}

\section{Introduction}\label{intro}

Among the more than 350 exoplanets known to date, transiting hot Jupiters present the first opportunity to study and understand the exoplanetary atmospheres.
Although they have masses similar to that of the giant planets from the solar system, they orbit extremely close to their host star (less than 0.1 AU) and we can expect them to be tidally locked 
to their parent star due to a fast tidal synchronisation. Their high equilibrium temperature ($1000-2000$~K) and the fact that they likely have permanent day/night sides presents interesting challenges and tests for planetary atmosphere models and atmospheric circulation.  



When the secondary eclipse of a transiting system occurs, it is possible to estimate the flux emitted by the day side of the planet relative to the star. The infrared \emph{Spitzer Space Telescope} has been used to detect the flux emitted by several exoplanets in its 6 photometric channels (3.6, 4.5, 5.8, 8.0, 16.0 and 24.0~\micron~bandpasses) and IRS spectrograph and tabulate the broadband infrared spectrum \citep{char05,char08,dem05,dem06,dem07,grill07,rich07,har07,demory07,knut08,mach08,grill08}. These observations haved pointed to the presence of two classes of hot Jupiters:

One class of planets, including HD 189733b \citep{dem06,grill07,char08,bar08} and TrES-1 \citep{char05}, have emission spectra consistent with standard 1D cloud-free atmosphere models for these planets \citep{hub03,sud03,seag05,bar05,fort05,fort06a,fort06b,burr05,burr06,burr08}. Their infrared spectra are dominated by absorption features from CO and H$_2$O.  

The other class, including HD 209458b \citep{dem05,rich07,burr07a,knut08}, TrES-2 \citep{odon09}, XO-1b \citep{mach08}, and TrES-4 \citep{knut09} have a temperature inversion between $0.1-0.01$ bars. Water bands that appear in emission instead of absorption \citep{fort06a,fort08,burr07b,burr08} are the most likely explanation of their observed spectrum. 


The first 6 hot Jupiters planets for which infrared measurements at two or more wavelengths have been presented have shown a connection between their equilibrium temperature and the presence of a temperature inversion.
It has been proposed \citep{hub03,burr07b,burr08,fort08} that these two different classes may be linked with TiO and VO molecules in the high atmosphere in a gas phase dependent on its effective temperature, that could lead to a temperature inversion from their opacity. However, the exoplanet XO-1b \citep{mach08} does not fit this rule, as it is shows evidence of a temperature inversion despite levels of irradiation comparable to those of HD 189733b and TrES-1.


With a period of only 31 hours (1.30619 days ; Sozzetti et al. 2009), TrES-3 (discovered by O'Donovan et al. 2006) has the shortest period of the known transiting exoplanets observable with Spitzer,
with the exception of the recently announced WASP-18b (Hellier et al. 2009). The very short period of TrES-3 results in a high level of irradiation, with an incident flux of $1.6$~$10^9 erg s^{-1} cm^{-2}$.
Its radius of $1.295~R_{Jup}$ is larger than predicted by simple models of the structure of these highly irradiated short period gas giants. The key to understanding the large radius may lie in the composition of the planetary atmosphere, which dictates the planet's cooling after formation and hence its final radius at its current age.
Several theoretical attempts have been made to propose an additional energy source in the planetary interior that would combat the planetary contraction after formation. 
Guillot et al. (2006) and \cite{burr07b} suggested that if the bloated planets have significantly enhanced metallicities, the resultant increased planetary opacity (and hence reduced contraction rate) could explain the large radii. Fressin et al. (2008) confirmed that an evolution model assuming both a linear correlation between the mass of the core (or heavy elements) of giant planets and their host star metallacity and an internal energy source was likely to reproduce quantitatively the distribution of masses and radii of the known transiting giant planets.

De Mooij \& Snellen (2009) have obtained for TrES-3 the first measurement of a planetary secondary eclipse depth from the ground, in $K$ band using the William Herschell telescope (WHT)
and the United Kingdom Infrared Telescope (UKIRT). They measured the $K$ band secondary eclipse depth of $0.241 \pm 0.043 \%$. 
This corresponds to a day-side brightness temperature at $2.2 \micron$ of $2040 \pm 185$ K in $K$ band. They also found the center of the secondary eclipse was slightly offset from orbital phase $\theta = 0.5$, of $\theta_0 = 0.5042 \pm 0.0027$, indicating that the orbit of TrES-3 was perhaps non-circular. If TrES-3 has a slightly eccentric orbit, tidal heating from ongoing circularization might provide enough energy to explain the planet's inflated radius. By measuring the timing of the secondary eclipse of TrES-3 in the 4 IRAC bandpasses, we will be able to constrain more precisely the planet's orbital eccentricity, either confirming or ruling out ongoing circularization as the explanation for the planet's inflated radius.

\section{Observations}\label{obs}

We used the Infrared Array Camera (IRAC ; \cite{faz04}) of the \emph{Spitzer Space Telescope} \citep{wern04} to observe the secondary eclipse of TrES-3 on UT 2008 July 18 and July 20, obtaining data at 3.6, 4.5, 5.8, and 8.0 \micron. We were able to observe it in full array mode in all 4 channels for a duration of 5.2 hours
We observed the target in the IRAC stellar mode, in which the camera gathers two 10.4 s integrations in the shorter wavelengths channels while gathering a single 30 s integration in the longer wavelengths channels. Therefore, we gathered 1248 images at  3.6, and 4.5 \micron and 624 images at 5.8 and 8.0 \micron.
We describe below our observations in two sections, as the InSb detectors used for IRAC channels at 3.6 and 4.5~\micron~requires a different treatement than the Si:As detectors of IRAC 5.8 and 8~\micron~channels. For these observations, we used the ``preflash'' technique (Knutson et al. 2009), in which we pointed the telescope towards a bright source before observing TrES-3. This was completed in order to reduce the amplitude of the detector ``ramp'' at 5.8, and 8.0 \micron, effectively pre-loading the pixels on which the target star would be pointed.

\subsection{3.6 and 4.5~\micron~Observations}\label{short_norm}

The contribution of the background to the total flux from TrES-3 is low in the 3.6 and 4.5~\micron~IRAC bandpasses, contributing only 0.3\% and 0.35\%, respectively, to the total flux in an aperture with a 5-pixel diameter centered on the position of the star. We obtain the lowest RMS time series using an aperture with a radius of 5.0 pixels. We allow the position of our aperture to shift with the position of the star in each image. 
We determine the position of the star in each image as the position-weighted sum of the flux in a 5-pixel radius disk centered on the approximate position of the star. We estimate the background in each image from an annulus with an inner radius of 12 pixels and an outer radius of 20 pixels centered on the position of the star. We calculate the JD value for each image as the time at mid-exposure and apply a correction to convert these JD values to the appropriate HJD, taking into account Spitzer's orbital position at each point during the observations.

\begin{figure}
\epsscale{1.1}
\plotone{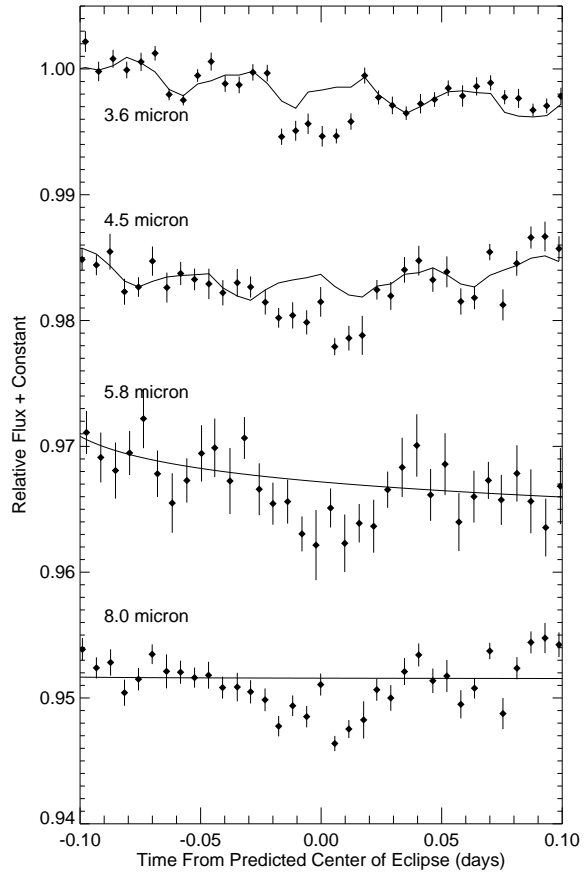}
\caption{Secondary eclipse of TrES-3 observed on UT 2008 July 20 at 3.6 and 5.8~\micron, and on UT 2008 Jul. 18 at 4.5 and 8~\micron.  Data are binned in 9.2 minute intervals and normalized to one, then offset by a constant for the purposes of this plot.  The overplotted curves show the best-fit corrections for detector effects (see \S\ref{short_norm} and \S\ref{long_norm}). \label{norm_plots}}
\end{figure}

\begin{figure}
\epsscale{1.1}
\plotone{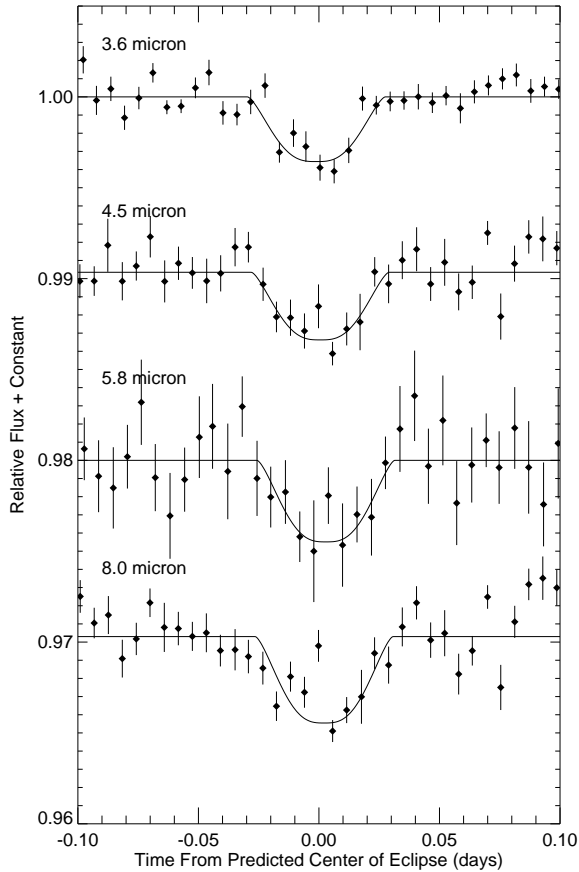}
\caption{Secondary eclipse of TrES-3 observed on UT 2008 Jul. 20 at 3.6 and 5.8~\micron, and on UT 2008 Jul. 18 at 4.5 and 8~\micron, with best-fit eclipse curves overplotted.  Data have been normalized to remove detector effects (see discussion in \S\ref{short_norm} and \S\ref{long_norm}), and binned in 9.2 minute intervals, then offset by a constant for the purposes of this plot.\label{four_eclipses}}
\end{figure}

The most important noise source in the first two IRAC bandpasses is due to a well-known intra-pixel sensitivity \citep{reach05,char05,char08,mor06,knut08}. Fluxes at these two wavelengths show a strong correlation with the intra-pixel position of the star on the detector, at a level comparable to the depth of the secondary eclipse. We use the following parameters to fit the observed flux as a linear function of the subpixel position:

\begin{equation}\label{eq1}
f=f_0*(c_1+c_2(x-x_0)+c_3(y-y_0))
\end{equation}

where $f_0$ is the original flux from the star, $f$ is the measured flux, $x$ and $y$ denote the location of the flux-weighted centroid of the star on the array, $x_0$ and $y_0$ are the coordinates of the center of the pixel containing the peak of the star's point spread function, and $c_1-c_3$ are free parameters in the fit.  In the 3.6~\micron~channel $x_0$ and $y_0$ had values of [171.5,175.5], and in the 4.5~\micron~channel they had values of [167.5,175.5]. 
We found that the position of the star on the array varied by 0.12 pixels in $x$ and 0.17 pixels in $y$ during our 3.6~\micron~observations. During our 4.5~\micron~observations the position of the star varied by 0.11 pixels in $x$ and 0.15 pixels in $y$. In contrast to previous observations of HD~189733 and HD~209458 in these channels \citep{knut08,char08}, we find that adding quadratic terms to this equation does not improve the fit, likely due to the lower SNR of the present observations. In both bandpasses the $\chi^2$ value for the fits is not improved by the addition of higher-order terms in $x$, $y$ and $x * y$, and the addition of these higher-order terms did not significantly alter the best-fit values for the best-fit eclipse times and depths.

After correcting for the intrapixel sensitivity, a trend is still visible at 3.6~\micron. 
Similar observations of TrES-4 (Knutson et al. 2008) a star with compatible brightness, also show the same kind of linear trends at 3.6~\micron~and 4.5~\micron, that is likely to be an instrumental effect related to the detector or telescope. We correct for this effect by fitting the data in both channels with a linear function of time.  This term is fitted simultaneously with the transit curve and the correction for the intrapixel sensitivity so that we can accurately characterize the additional uncertainty in the depth and timing of the eclipse introduced by these corrections.  This means that at 3.6~\micron~we are fitting for six parameters, including a constant term, a linear function of $x$ position, a linear function of $y$ position, a linear function of time, the eclipse depth, and the eclipse time. 
We fit the data using a Markov Chain Monte Carlo method \citep{ford05,winn07} with $10^6$ steps, where we set the uncertainty on individual points equal to the standard deviation of the out-of-transit data after correction for the various detector effects. 


Before beginning our fit we do an initial trim within our aperture, discarding outliers higher than 3.5~$\sigma$ of the local median flux (defined as the median of a 15 minutes window centered on the data point). We also remove measurements for which the identified position of the photocenter $x$ or $y$ deviates more than 3.5~$\sigma$ from the same 15-minutes median position. This global triming respectivly excludes 6 and 7\% of the data points in the 3.6 and 4.5~\micron~bandpasses. 

Next we carry out the Markov chain fit on the trimmed data. We allow both the depth and time of the secondary eclipse to vary independently for the eclipses at each of the two observed wavelengths, and take the other parameters for the system (planetary and stellar radii, orbital period, semi-major axis and inclination) from \citet{sozz09}.  We calculate our eclipse curve using the equations from \citet{mand02}. 
During each step of the chain we exclude outliers greater than either 3.5$\sigma$~(for both the 3.6 and 4.5~\micron~fits), as determined using the residuals from the model light curve, from our evaluation of the $\chi^2$ function.  We rescale the value of the $\chi^2$ function to account for the fact that we are varying the number of pixels included in the fit. 

After running the chain, we search for the point in the chain where the $\chi^2$ value first falls below the median of all the $\chi^2$ values in the chain (i.e. where the code had first found the best-fit solution), and discard all the steps up to that point.  We take the median of the remaining distribution as our best-fit parameter, with errors calculated as the symmetric range about the median containing 68\% of the points in the distribution.  The distribution of values is very close to symmetric and gaussian for the 5 parameters we fitted together ($c_1,c_2,c_3$ and transit depth and time), and we checked that there were no strong correlations between variables. 
Table \ref{eclipse_depths} states our results for the eclipses depths and times and we plot the time series in Fig 1. and Fig 2.

\subsection{5.8 and 8.0~\micron~Observations}\label{long_norm}

Previous secondary eclipse studies (e.g. Knutson et al. 2008) have shown that PSF-fitting can provide a better signal-to-noise ratio at longer wavelengths. At longer wavelengths the flux from the star is smaller and the zodiacal background is larger; we find that the background contributes 14\% and 16\% of the total flux in a 3.-pixel aperture at 5.8 and 8.0~\micron, respectively. Because the background is higher in these two channels (the median background flux is 1.2 MJy/Sr in the 5.8~\micron~bandpass and 0.6 MJy/Sr in the 8.0~\micron~bandpass), we used a PSF fit to derive the time series in both bandpasses and compared the results to those from aperture photometry. 

At 5.8~\micron~and 8.0~\micron~respectively, we found that the relative scatter in the time series after model fitting from the PSF fits was 20\% and 25\% higher than in the time series from aperture photometry with a 3.0 pixel radius.  As a result of this increased scatter, which is likely produced by discrepancies between the model PSF and the observed PSF, we conclude that aperture photometry is also preferable in these 2 channels. We compare the time series using apertures ranging from $3-4.5$ pixels and find consistent results in all cases, but with a scatter that increases with the radius of the photometric aperture.



\begin{deluxetable*}{lrrrrcrrrrr}
\tabletypesize{\scriptsize}
\tablecaption{Best-Fit Eclipse Depths and Times \label{eclipse_depths}}
\tablewidth{0pt}
\tablehead{
\colhead{$\lambda$ (\micron)} & \colhead{Eclipse Depth}  & \colhead{Center of Transit (HJD)} & \colhead{O$-$C (min.)\tablenotemark{a}}}
\startdata
3.6 & $0.356\pm0.035\%$ & $2454668.5447\pm0.0020$ & $-2.0\pm2.9$\phantom{0}\\
4.5 & $0.372\pm0.054\%$ & $2454665.9343\pm0.0027$ & $0.9\pm3.9$\phantom{0}\\
5.8 & $0.449\pm0.097\%$ & $2454668.5498\pm0.0042$ & $5.4\pm6.0$\phantom{0}\\
8.0 & $0.475\pm0.046\%$ & $2454665.9365\pm0.0021$ & $4.0\pm3.0$\phantom{0}\\
\enddata
\tablenotetext{a}{Observed minus calculated transit times, where the expected transit times are calculated using the ephemeris from \citet{mand07} and assuming zero eccentricity.}
\end{deluxetable*}

\begin{figure}
\epsscale{1.1}
\plotone{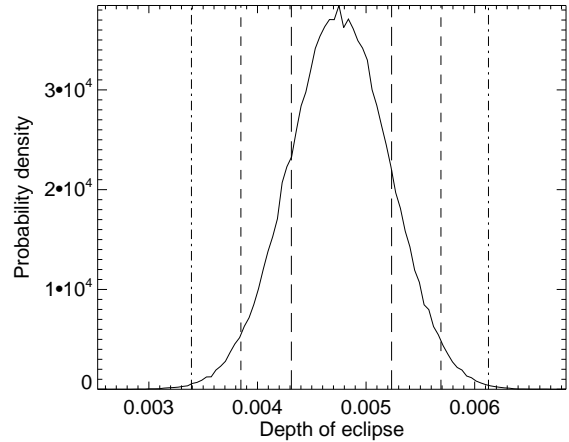}
\caption{Probability distribution for the eclipse depth from a Markov Chain Monte Carlo fit to the IRAC 8~\micron~ data. 
The long-dashed, short-dashed and dot-dashed a lines indicate the $1\sigma$, $2\sigma$ and $3\sigma$ limits on the eclipse depth, respectively, which were calculated by integrating over this distribution.\label{hist}}
\end{figure}

The 5.8 and 8~\micron~bandpasses of IRAC camera are known to be affected by a ``detector ramp'' \citep{knut08,knut09,char08} that causes the effective gain (and thus the measured flux) in individual pixels to increase. 
The size of this effect depends on the illumination level of the individual pixel. Pixels with high illumination ($>~250 MJy/Sr$ in the 8~\micron~channel) will converge to a constant value within the first hour of observations, whereas lower-illumination pixels will show a linear increase in the measured flux over time with a slope that varies inversely with the logarithm of the illumination level. 
Preflashing the detector array by pointing a bright source prior to conducting the science observations can reduce the amplitude of this ramp by doing a pre-load of the pixels on which the target star will be pointed. In our observation of TrES-3, we targeted a bright star as a preflash source and this yielded one of the smallest ramp effect ever recorded using 5.8 and 8~\micron~observations. 
Previous observations (e.g. \citet{knut09}) have shown that the ramp is well described as following an asymptotic shape, with a steeper rise in the first 30 minutes of observations. We correct for this effect by fitting our time series in both bandpasses with the following function:

\begin{equation}\label{eq2}
f=f_0*(c_1+c_2 ln(dt)) 
\end{equation}

where $f_0$ is the original flux from the star, $f$ is the measured flux, and $dt$ is the elapsed time in days since the start of the observations. 
We fit both Eq. \ref{eq2} and the transit curve to the data simultaneously using a Markov Chain Monte Carlo method as described in \S\ref{short_norm}. As before, the distribution of values was very close to symmetric in all cases, and there were no strong correlations between the variables.  Best-fit eclipse depths and times from these fits are given in Table \ref{eclipse_depths}, and the time series before and after correcting for detector effects are shown in Figures \ref{norm_plots} and \ref{four_eclipses}, respectively. As a check we repeated these fits adding a quadratic term of $ln(dt)$ in Eq. \ref{eq2}, and found that the value of the $\chi^2$ function for our best-fit solution were similar at both $5.8$ and $8.0$~\micron~to their previous values. In order to estimate error bars for the measured depth and time, we take the median of the remaining distribution as our best-fit parameter, with errors calculated as the symmetric range about the median containing 68\% of the points in the distribution.  Figure \ref{hist} shows the representative example of the histogram of the probability distribution for the eclipse depth at 8.0 ~\micron. 

\begin{figure*}
\epsscale{1.1}
\plotone{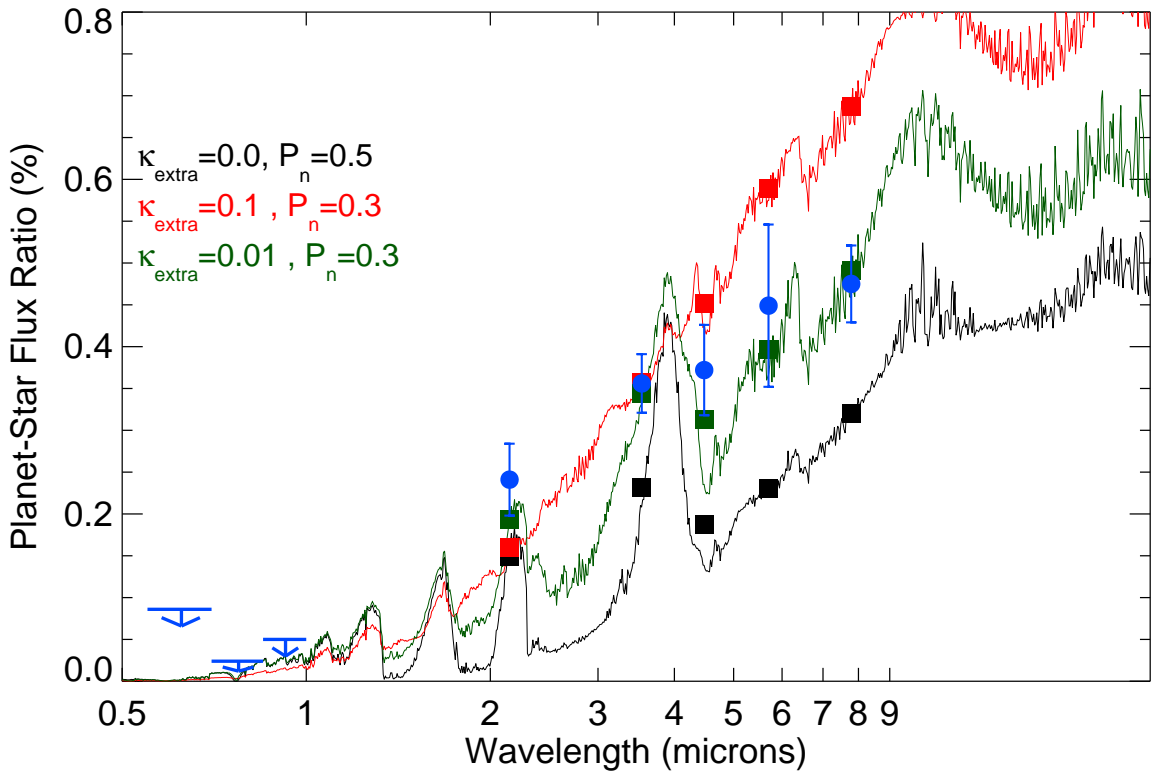}
\caption{Day-side planet-star flux ratios for TrES-3 as determined from measurements of the secondary eclipse depth in the four IRAC bandpasses (blue circles) and in the $K$ band UKIRT detection. We also show the three 99~\%-confidence upper limits obtained by Winn et al. 2008 in the $R$, $i$, and $z$ bands.
The black line corresponds to a default model with no temperature inversion with a redistribution parameter $P_n=0.5$, which describes the case where the incident energy is fully redistributed across the entire surface of the planet. The red and green lines correspond to models with an additional optical absorber at high altitudes (parameterized as $\kappa_{extra}$), which produces a thermal inversion around pressures of 0.001 bar \citep{burr07a,burr08}. Squares show the values for these models after integrating over the \emph{Spitzer} bandpasses. 
The eclipse depths in the 3.6, 4.5 and 5.8~\micron~bandpasses can be matched by both models with relatively efficient day-night circulation $P_n=0.3$ and different levels of additional opacity (red model $\kappa_{extra}=0.1$ cm$^2$/g and green model $\kappa_{extra}=0.01$). However, the planet-star flux ratio at 8.0~\micron~argues strongly against the presence of a temperature inversion, as the green model with a low $\kappa_{extra}=0.01$ cm$^2$/g provides the best match at this wavelength ; the K-band UKIRT detection also slightly favors this scenario.  \label{spectrum}}
\end{figure*}

\section{Discussion}


We determine the best-fit eclipse times for the four secondary eclipses observed using IRAC by taking the weighted average of the best-fit eclipse times in each bandpass. Using this method, we find that the eclipse is shifted by $1.0 _{-1.3} ^{+1.9}$ minutes later than the expected time based on the ephemeris from Sozzetti et al. (2009). 

Our estimate for the best-fit timing offset translates to a constraint on the orbital eccentricity $e$ and the argument of pericenter $\omega$ of $ecos(\omega)=0.00084_{-0.0009}^{+0.0016}$; the $3\sigma$ upper limit on its absolute value is $|ecos(\omega)|<0.0056$. We selected these limits because we are interested in constraining the magnitude of $e$ rather than the sign of the $cos(\omega)$ term.  This upper limit means that unless the longitude of periastron $\omega$ is close to 90\degr~or 270\degr, we can rule out tidal heating from ongoing orbital circularization \citep{bod01,liu08} as an explanation for the inflated radius of TrES-3. Winn et al. (2008) have placed a constraint on the albedo assuming a circularized orbit. Thus, our results serve to validate this assumption, and we proceed to adopt their stated upper limits in our analysis. They placed a $99\%$-confidence upper limits on the planet-to-star flux ratio of $2.4 \times 10^{-4}$, $5.0 \times 10^{-4}$, and $8.6 \times 10^{-4}$ in the $i$, $z$, and $R$ bands respectively.



\begin{figure}
\epsscale{1.1}
\plotone{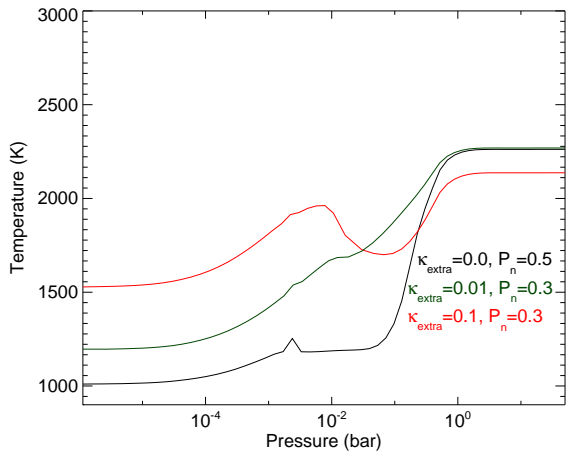}
\caption{Day-side pressure-temperature profiles for the three models plotted in Fig. \ref{spectrum}.  The temperature increases at low pressures as a function of the amount of the opacity of
the $\kappa_{extra}$ absorber. The fraction of energy redistributed to the night side of the planet $P_n$ decreases the temperature at lower atmospheric levels (0.01 - 0.1 bars). 
A drop in temperature (described in Burrows et al. 2007b, 2008) occurs as the day-night circulation is turned up.
\label{ptplot}}
\end{figure}

We then compare the secondary eclipse depths in the four IRAC bandpasses and the previously reported $K$ band value to the predictions from atmosphere models for this planet (see Fig. \ref{spectrum} and \ref{ptplot}). We employ the formalism described in \citet{burr07a,burr08}. We use a stellar atmosphere model \citep{kurucz79,kurucz94,kurucz05} with an effective temperature of $5650$~K and a planet-star radius ratio of $0.1654$ based on the measures of \citet{sozz09}. 
We calculate the emergent spectrum at secondary eclipse for a pair of free parameters, P$_n$ and $\kappa_{extra}$. P$_n$ is the dimensionless redistribution parameter that accounts for the cooling of the dayside and the warming of the nightside by zonal winds near an optical depth of order unity, ranging form $0$ to $0.5$. It is a measure of the efficiency of heat redistribution by super-rotational hydrodynamic flows. As the value of P$_n$ is increased, the day side becomes cooler and the emergent planetary flux at superior conjunction becomes correspondingly small. $\kappa_{extra}$ is the absorptive opacity in the optical at altitude (here in cm$^2$/g) and our best fit requires only a very low opacity that does not create a temperature inversion. 

\begin{deluxetable*}{cccc}
\tabletypesize{\scriptsize}
\tablecaption{$\chi^2$ values for different atmospheric models \label{chi_squares}}
\tablewidth{0pt}
\tablehead{
\colhead{$\kappa_{extra}$} & \colhead{$P_n$}  & \colhead{$\chi^2$} & \colhead{$\chi^2/n$}\tablenotemark{a}}
\startdata
$0.00$ & $0.1$ & $1.317$ & $1.105$ \\
$0.01$ & $0.1$ & $3.895$ & $1.289$ \\
$0.1$  & $0.1$ & $41.10$ & $13.70$ \\
$0.00$ & $0.5$ & $40.77$ & $13.59$ \\
$0.01$ & $0.5$ & $24.60$ & $8.210$ \\
$0.1$  & $0.5$ & $38.43$ & $12.76$ \\
$0.00$ & $0.3$ & $6.346$ & $2.115$ \\
$0.01$ & $0.3$ & $1.734$ & $0.578$ \\
$0.025$& $0.3$ & $2.645$ & $0.881$ \\
$0.05$ & $0.3$ & $9.400$ & $3.132$ \\
$0.1$  & $0.3$ & $25.53$ & $8.511$ \\
\enddata
\tablenotetext{a}{$\chi^2$/n is the reduced $\chi^2$ with n $= 5$ parameters $- 2$ degrees of freedom.}

\end{deluxetable*}

Fig. \ref{spectrum} and \ref{ptplot} show three models with different values for $P_n$ and $\kappa_{extra}$. The standard non-inverted model ($\kappa_{extra}=0$ cm$^2$/g) is clearly inconsistent with the observed fluxes from TrES-3 at wavelengths longer than 3~\micron. It is possible to match the observed 3.6~\micron~flux with this model by reducing the relative fraction of the incident energy that is redistributed to the planet's night side, thus increasing the day-side temperature and corresponding fluxes. 
We plotted the best-fit model, that involves no thermal inversion ($\kappa_{extra}=0.01$ cm$^2$/g), and an example of an inverted ($\kappa_{extra}=0.1$ cm$^2$/g) model. Although, for each of the four individual planet-star flux ratios we are able to find a model with a temperature inversion that fits well, we are unable to find a single model that fits all four data points simultaneously. In the ($\kappa_{extra}=0.1$ cm$^2$/g) scenario involving a temperature inversion, the 8.0~\micron~flux can not be reproduced well, and we concluded that the model with no inversion provides an overall better fit. The $K$ band measurement also strengthens the conclusion that the model with no inversion provides a better global fit to the data. We compared the $\chi^2$ values for 5 data points and 2 free parameters for several models including the three plotted models: table \ref{chi_squares} shows the $\chi^2$ values we obtain these different models. Only non-inverted models provide a good fit to the 5 data points, with the plotted model ($\kappa_{extra}=0.01$ cm$^2$/g and $P_n=0.3$) showing the best solution.

\section{Conclusions}

We have detected the TrES-3 secondary eclipse in the 4 bandpasses of the IRAC instrument. These observations at 3.6, 4.5, 5.8, and 8.0~\micron, combined with the K band measuement from De Mooij \& Snellen (2009), reveal that this planet does not show a thermal inversion similar to the one observed for HD 209458b \citep{knut08,burr07a}, TrES-2 \citep{odon09} and TrES-4 \citep{knut09}. 
The best overall fit involves an efficient day-night circulation ($P_n=0.3$) and a very low additional opacity ($\kappa_{extra}=0.01$ cm$^2$/g). The scenarios presented by Fortney et al. (2008) would predict that gas phase TiO or VO at high altitude would result in a temperature inversion for this highly-irradiated temperature planet, as it is warmer than HD209458, TrES-2 and TrES-4, which are inverted. The fact that our results strongly favor a scenario without any temperature inversion shows that the distinction is not simply due to the level of irradiation for the separation between these two kinds of close-in giant planets. Spiegel et al. (2009) have shown that the TiO-VO hypothesis was unlikely unless there is significant mixing in the atmosphere, and TrES-3 could be among the planets were the mixing is insufficient.

With an upper limit of $|ecos(\omega)|<0.0056$ for the orbital eccentricity, we can rule out tidal heating from ongoing orbital circularization at the level required by \citet{liu08} in order to explain TrES-3's inflated radius. This is also the first exoplanet in this range of high irradiation level not to show a temperature inversion.
This interesting and unexpected case emphasizes the importance of gathering more hot Jupiter infrared emission measurements, in order to study the correlations between temperature inversion and system characteristics.
A large sample of multi-wavelength infrared measurements from many different exoplanets will be required to understand the origin of these temperature inversions. Fortunately, this will be accessible during the Warm phase of the \emph{Spitzer} mission, as both the 3.6 and 4.5~\micron~channels will continue to function at full sensitivity. Thirty-four known transiting exoplanets known to date are bright enough for the two \emph{Spitzer} observations at 3.6 and 4.5 \micron~to assess or rule out if a temperature inversion occurs in their high atmosphere. 
Observations of the secondary eclipse in these two channels should be sufficient to distinguish between planets with and without temperature inversions in most cases, although the case of TrES-3 shows that the additional information at longer wavelengths can be the discriminating factor between these two possible scenarios. The ground-based high precision detection of thermal emission is also helpful.
The global set of observations, combining detections in the 4 IRAC bandpasses obtained during the cryogenic \emph{Spitzer} mission for 15 planets, and the observations of 19 other planets during the warm phase of the mission by the Spitzer Exploratory Science Program 60021 (Principal Investogator H. Knutson) will allow the study of correlations of the temperature inversion phenomenon with stellar metallicities, planet size and mass, levels of irradiation, surface gravities, and orbital periods. Studying the connection between the presence of a temperature inversion and these characteristics will give us a better understanding of planetary atmospheres under strong irradiation.

\acknowledgements

This work is based on observations made with the \emph{Spitzer Space Telescope}, which is operated by the Jet Propulsion Laboratory, California Institute of Technology, under contract to NASA. 

\end{document}